\documentstyle[amssymb,aps]{revtex}

\begin{document}
\title{Dynamics of the Magnetic Flux Trapped in Fractal Clusters of a Normal Phase
in Superconductor}
\author{Yu. I. Kuzmin}
\address{Ioffe Physical Technical Institute of Russian Academy of Sciences,\\
Polytechnicheskaya 26 St., Saint Petersburg 194021 Russia\\
e-mail: yurk@shuv.ioffe.rssi.ru; iourk@usa.net\\
tel.: +7 812 2479902; fax: +7 812 2471017}
\date{\today}
\maketitle

\begin{abstract}
The influence of geometry and morphology of superconducting structure on
critical currents and magnetic flux trapping in percolative type-II
superconductor is considered. The superconductor contains the clusters of a
normal phase, which act as pinning centers. It is found that such clusters
have significant fractal properties. The main features of these clusters are
studied in detail: the cluster statistics is analyzed; the fractal dimension
of their boundary is estimated; the distribution of critical currents is
obtained, and its peculiarities are explored. It is examined thoroughly how
the finite resolution capacity of the cluster geometrical size measurement
affects the estimated value of fractal dimension. The effect of fractal
properties of the normal phase clusters on the electric field arising from
magnetic flux motion is investigated in the case of an exponential
distribution of cluster areas. The voltage-current characteristics of
superconductors in the resistive state for an arbitrary fractal dimension
are obtained. It is revealed that the fractality of the boundaries of the
normal phase clusters intensifies the magnetic flux trapping and thereby
raises the critical current of a superconductor.
\end{abstract}

\section{INTRODUCTION}

An important property of the clusters of a normal phase in superconductor
consists in their capability to trap a magnetic flux. By virtue of their
capacity to hold the vortices from moving under the action of the Lorentz
force, such clusters can act as effective pinning centers \cite{h1}-\cite{s5}%
. This feature is used widely in the making new composite superconducting
materials of high current-carrying capability \cite{m6}, \cite{b7}. The
morphological characteristics of clusters of a normal phase exert an
appreciable effect on magnetic flux dynamics in superconductors, especially
when the clusters have fractal boundaries \cite{l8}-\cite{k10}. In the
present work the geometric probability properties of such fractal clusters
are considered in detail, and their influence on the dynamics of trapped
magnetic flux and critical currents is analyzed.

The further consideration will be concerned with the superconductor
containing inclusions of a normal phase, which are out of contact with one
another. Let us assume that these inclusions are oriented in such a way that
their extent along one of directions far exceeds other linear sizes. The
similar columnar defects are of most interest for creating the artificial
pinning centers \cite{m6}, \cite{k10}-\cite{k14}. When such a
superconducting structure is cooled below the critical temperature in the
magnetic field along the direction of the longest size of these inclusions,
the magnetic flux will be frozen in the normal phase clusters. Even after
the external field has been turned off, the flux trapped in these clusters
is kept unchanged due to the currents that are steadily circulating around
them through the superconducting loops. The distribution of the trapped
magnetic flux resulting from such a magnetization in the field-cooling
regime will be two-dimensional. The similar distribution can readily be
realized in the superconducting film where the normal phase inclusions are
created during the growth process at the sites of defects on the boundary
with the substrate in such a way that their orientation is normal to the
surface of the film \cite{m6}, \cite{k13}, \cite{k14}. Let us suppose that
the film surface fraction covered by the normal phase is below the
percolation threshold for the transfer of magnetic flux (50 \% for 2D-
percolation \cite{s15}). In this case the relative portion of
superconducting phase exceeds the percolation threshold, so there is a
superconducting percolation cluster in the plane of the film where a
transport current can flow. Such a structure provides for effective pinning
and thereby raises the critical current, because the magnetic flux is locked
in finite clusters of a normal phase, and so the vortices cannot leave them
without crossing the surrounding superconducting space. If the transport
current is passed through the sample, the trapped magnetic flux remains
unchanged as long as the vortices are still held in the normal phase
clusters. When the current is increased, the magnetic flux starts to break
away from the clusters of pinning force weaker than the Lorentz force
created by the transport current. As this takes place, the vortices will
first pass through the weak links, which connect the normal phase clusters
between themselves. Such weak links form especially readily in
high-temperature superconductors (HTS) characterized by an extremely short
coherence length. Various structural defects, which would simply cause some
additional scattering at long coherence length, give rise to the weak links
in HTS. There is a hierarchy of weak links over a wide range of scales in
HTS \cite{m6}, \cite{b16}-\cite{h24}. At an atomic level the weak links are
formed by the structural atomic defects, primarily, by oxygen vacancies \cite
{s18}, \cite{r25}. On a mesoscopic scale twin boundaries are mainly
responsible for weak links existence \cite{s18}-\cite{k21}, \cite{m26}. The
twins can be spaced up to several nanometers apart, so even single crystal
has fine substructure caused by twins. At last, on a macroscopic scale there
are manifold structural defects which can form weak links: that may be grain
or crystallite boundaries as well as barriers arising from the secondary
degrading the non-stoichiometric crystal into the domains with a high and
low content of oxygen \cite{k22}-\cite{h24}. Moreover, a magnetic field
further reduces a coherence length \cite{s27}, thus resulting in more easy
weak links formation. In conventional low-temperature superconductors, which
are characterized by a large coherence length, the weak links can be formed
due to the proximity effect in sites of minimum distance between the next
normal phase clusters.

As soon as the transport current is turned on, this one is added to all the
persistent currents, which maintain the magnetic flux trapped. Each of these
currents is circulating through the superconducting loop around the normal
phase cluster wherein the corresponding portion of the magnetic flux is
trapped. The loop contains weak links that join the adjacent normal phase
clusters transversely to the path of the current. As the transport current
is increased, there will come a point when the overall current flowing
through the weak link will exceed the critical value, so this link will turn
into a resistive state. As this takes place, the space distribution of the
currents throughout the superconducting cluster is changed in such a way
that the resistive subcircuit will be shunted by the superconducting paths
where weak links are not damaged yet. Magnetic field created by this
re-distributed transport current acts via the Lorentz force on the current
circulating around the normal phase cluster. As a consequence, the magnetic
flux trapped therein will be forced out through the resistive weak link,
which has become permeable to the vortices.

\section{FRACTAL GEOMETRY OF NORMAL PHASE CLUSTERS}

Thus, whatever the microscopic nature of weak links may be, they form the
channels for vortex transport where the pinning force is weaker than the
Lorentz force created by the transport current. It appears that according to
their configuration each normal phase cluster has its own value of the
critical current of depinning, which contributes to the overall statistical
distribution. When a transport current is gradually increased, the vortices
will break away first from clusters of small pinning force, and therefore,
of small critical current. Thus the decrease in the trapped magnetic flux $%
\Delta \Phi $ is proportional to the number of all the normal phase clusters
of critical currents less than a preset value $I$. Therefore, the relative
decrease in the trapped flux can be expressed with the cumulative
probability function $F=F(I)$ for the distribution of the critical currents
of clusters: 
\begin{equation}
\frac{\Delta \Phi }{\Phi }=F(I)\text{ \ \ \ \ \ \ \ , \ \ \ \ \ \ where \ \
\ \ \ \ \ \ \ \ \ \ \ \ \ \ \ \ \ }F(I)=\Pr \left\{ \forall I_{j}<I\right\}
\label{probi1}
\end{equation}
The right-hand side of Eq.~(\ref{probi1}) is the probability that any $j$-th
cluster has the critical current $I_{j}$ less than a given upper bound $I$.

On the other hand, the magnetic flux trapped into a single cluster is
proportional to its area $A$, so the decrease in the total trapped flux can
be represented by the cumulative probability function $W=W(A)$ for the
distribution of the areas of the normal phase clusters, which is a measure
of the number of the clusters of area smaller than a given value of $A$: 
\begin{equation}
\frac{\Delta \Phi }{\Phi }=1-W(A)\text{ \ \ \ \ \ , \ \ \ \ \ \ where \ \ \
\ \ \ \ \ \ \ \ \ \ \ \ \ \ }W(A)=\Pr \left\{ \forall A_{j}<A\right\}
\label{proba2}
\end{equation}

The distribution function $W=W(A)$ of the cluster areas can be found by the
geometric probability analysis of electron photomicrographs of
superconducting films \cite{k13}. So, in the practically important case of
thick YBCO films containing columnar defects \cite{k13}, \cite{k14} the
exponential distribution can be realized: 
\begin{equation}
W(A)=1-\exp \left( -\frac{A}{\overline{A}}\right)  \label{expa3}
\end{equation}
where $\overline{A}$ is the mean area of the cluster.

Thus, in order to clear up how the transport current acts on the trapped
magnetic flux, it is necessary to find out the relationship between the
distribution of the critical currents of the clusters (Eq.~(\ref{probi1}))
and the distribution of their areas (Eq.~(\ref{proba2})). The larger the
cluster size, the more weak links rise from its perimeter bordering with the
surrounding superconducting space, and therefore, the smaller is the
critical current at which the magnetic flux breaks away from this cluster.
On the basis of this simple geometric argument, let us suppose that this
critical current $I$ is inversely proportional to the perimeter $P$ of the
normal phase cluster: $I\propto 1/P$, because the larger the perimeter is,
the higher is the probability to get a weak link there. At this point it is
assumed that the weak link concentration per unit perimeter length is
constant for all clusters and all the clusters of equal perimeter have the
same pinning force. Also suppose that after the vortex enters the weak link,
it will pass all the way between two adjacent normal phase clusters without
being trapped. Here, the magnetic flux is transferred by Josephson vortices.
The Josephson penetration depth is large enough in the considered materials,
so the size of the region, where the vortex is localized, much exceeds the
characteristic length of all possible structural defects that can occur
along the transport channel. Thus the probability that such a vortex will be
trapped in passing through a weak link under the action of the Lorentz force
is very small.

Thus, to deal with the distribution function of Eq.~(\ref{probi1}), the
relation between perimeter and area of clusters should be studied. As has
been first found in Ref.~\cite{k10}, the fractal nature of the normal phase
clusters exerts an appreciable effect on the dynamics of a magnetic flux in
superconductors. For fractal clusters a relation between perimeter and area
has the form: 
\begin{equation}
P\propto A^{\frac{D}{2}}  \label{scaling4}
\end{equation}
where $D$ is the fractal dimension of the cluster perimeter (so-called
coastline dimension) \cite{m28}.

The relation of Eq.~(\ref{scaling4}) is consistent with the generalized
Euclid theorem \cite{m28}, \cite{m29}, which states that the ratios of the
corresponding measures are equal when reduced to the same dimension. Hence
it follows that $P^{1/D}\propto A^{1/2}$, which is valid both for Euclidean (%
$D=1$), and for fractal clusters ($D>1$).

The fractal dimension value can be estimated by means of regression analysis
of the sampling of the areas and perimeters of the normal phase clusters.
Such geometric probability analysis was carried out by the procedure
described in Ref.~\cite{k10}. For this purpose an electron photomicrograph
of YBCO film, which was similar to that published earlier in Ref.~\cite{k14}%
, has been scanned. The perimeters and areas of clusters have been measured
by covering their digitized pictures with a square grid of spacing $60\times
60\,$nm$^{2}$. The results of the statistical treatment of these data are
presented in Table~\ref{table1}. The normal phase has occupied 20\% of the
total surface only, so the transport current can flow through the
sufficiently dense percolation superconducting cluster. The primary sampling
has contained 528 normal phase clusters located on the scanned region of a
total area of 200~$\mu $m$^{2}$. The distribution of the cluster areas is
fitted well to exponential cumulative probability function of Eq.~(\ref
{expa3}) with the mean cluster area $\overline{A}=0.0765$~$\mu $m$^{2}$. All
the points of the primary sampling are marked by crosses on the plot~1 of
Fig.~\ref{figure1}, which shows the perimeter-area relation for the normal
phase clusters. Figure~\ref{figure1} also demonstrates such an important
peculiarity of this relation: the scaling law of Eq.~(\ref{scaling4}) is
valid in the range of almost three orders of magnitude in cluster area. The
scaling perimeter-area behavior means that there is no characteristic length
scale between 0.1~$\mu $m and 10~$\mu $m in the linear size of the normal
phase cluster. Whatever the shape and size of the clusters may be, all the
points fall closely on the same straight line in logarithmic scale; so that
there are no apparent kinks or crossovers on the graph. This enables the
fractal dimension $D$ of the cluster perimeter to be estimated from the
slope of the regression line of the form of Eq.~(\ref{scaling4}), thus a
least squares treatment of the perimeter-area data for the primary sampling
gives the estimate of $D=1.44\pm 0.02$ with correlation coefficient 0.929.

This point that the found value of coastline fractal dimension differs
appreciably from unity engages a great attention. What this means is that
the fractal properties of the cluster boundary are of prime importance here.
Two straight lines (5) in Fig.~\ref{figure1} bound the range of the slopes
that the dependencies of the perimeter on the cluster area can have for any
arbitrary fractal dimension. The least slope corresponds to Euclidean
clusters $(D=1)$, the most one relates to the clusters of the greatest
possible coastline dimension, which is equal to the topological dimension of
a smooth surface $(D=2)$. Such a fractal dimension is inherent, for example,
in Peano curves, which fill the whole plane \cite{m28}. Whatever the
geometric morphological properties of clusters may be, the slope of their
perimeter-area graphs will be always bounded by these two limiting lines.

When we deal with the geometric features of the normal phase clusters, we
are considering the cross section of the extended columnar defects by the
plane carrying a transport current. Therefore, though normal phase clusters
are self-affine fractals [30, 31], it is possible to examine their geometric
probability properties in the planar section only, where the boundaries of
the clusters are statistically self-similar.

Next, using the relation of Eq.~(\ref{scaling4}) between the fractal
perimeter and the area of the cluster, as well as our initial assumption
that the critical current of each cluster is inversely proportional to its
perimeter, we get the following expression for the critical current: $%
I=\alpha A^{-D/2}$, where $\alpha $ is the form factor. In accordance with
starting formulas of Eq.~(\ref{probi1}) and Eq.~(\ref{proba2}), the
exponential distribution of cluster areas of Eq.~(\ref{expa3}) gives rise to
an exponential-hyperbolic distribution of critical currents: 
\begin{equation}
F(i)=\exp \left[ -\left( \frac{2+D}{2}\right) ^{\frac{2}{D}+1}i^{-\frac{2}{D}%
}\right]  \label{exhyp5}
\end{equation}
where $i\equiv I/I_{c}$ is the dimensionless transport current, and $%
I_{c}=\left( \frac{2}{2+D}\right) ^{\frac{2+D}{2}}\alpha \left( \overline{A}%
\right) ^{-\frac{D}{2}}$ is the critical current of the resistive transition.

Thus, the geometric probability properties of the normal phase clusters are
responsible for main features of the critical current statistical
distribution. In turn, given the distribution of Eq.~(\ref{exhyp5}), the
change in the trapped magnetic flux caused by the transport current can be
found with the aid of Eq.~(\ref{probi1}). Two of these graphs are displayed
in Fig.~\ref{figure2} both for the fractal clusters of coastline dimension $%
D=1.44$ found above (curve (2)) and for the Euclidean ones (curve (3)).

In order to get the relationship between the dynamics of the trapped
magnetic flux and geometric morphological properties of the superconducting
structure, the sample empirical function $F^{\ast }=F^{\ast }(i)$ of the
distribution of the critical currents has been computed. This function gives
a statistical estimate of the cumulative probability function $F=F(i)$.
First, the empirical distribution function $W^{\ast }=W^{\ast }(A)$ for the
primary sampling of the areas of the normal phase clusters has been found.
The value $W^{\ast }\left( A\right) $ was calculated for each order
statistic as the relative number of clusters of area smaller than a given
value $A$. Next, the empirical distribution of the critical currents was
computed for the same order statistics (step line with open circles (1) in
Fig.~\ref{figure2}) using the following transformations: 
\[
\left. 
\begin{array}{c}
F^{\ast }=1-W^{\ast } \\ 
i=\left( \frac{2+D}{2}\right) ^{\frac{2+D}{2}}\left( \frac{\overline{A}}{A}%
\right) ^{\frac{D}{2}}
\end{array}
\right\} 
\]

Figure \ref{figure2} shows that in the range of currents $i<6$ the empirical
distribution function, which describes the morphological properties of the
superconducting structure (plot (1)), coincides ideally with the cumulative
probability function (curve (2)) for the coastline fractal dimension of $%
D=1.44$. Starting with the value of the current $i=6$ the crossover is
observed, resulting in that the empirical distribution function passes to
the dependence for Euclidean clusters $(D=1)$. This transition into the
Euclidean region is over at large transport currents, when the magnetic flux
changes mainly for the breaking of the vortices away from the small clusters
(as the smaller clusters have the larger pinning force). The observed
crossover has its origin in the finite resolution capability of measuring
the cluster geometrical sizes. The distinctive feature of the topologically
one-dimensional fractal curve is that its measured length $P$ depends on the
measurement accuracy in such a way: $P\propto \delta ^{1-D}$, where $\delta $
is the yardstick size used to measure this length, $(1-D)$ is the Hausdorff
codimension for the Euclidean 1D-space \cite{m28}. In our case such a
fractal curve is represented by the boundary of the normal phase cluster.
That is why just the statistical distribution of the cluster areas, rather
than their perimeters, is fundamental for finding the critical current
distribution of Eq.~(\ref{exhyp5}). The topological dimension of perimeter
is equal to unity and does not coincide with its Hausdorff-Besicovitch
dimension, which strictly exceeds the unity. Therefore the perimeter length
of a fractal cluster is not well defined, because its value diverges as the
yardstick size is reduced infinitely. On the other hand, the topological
dimension of the cluster area is the same as the Hausdorff-Besicovitch one
(both are equal to two). Thus, the area restricted by the fractal curve is
well-defined finite quantity.

Taking into account the effect of the measurement accuracy, the
perimeter-area relationship of Eq.~(\ref{scaling4}) can be re-written as 
\begin{equation}
P(\delta )\propto \delta ^{1-D}\left( A\left( \delta \right) \right) ^{\frac{%
D}{2}}  \label{perima6}
\end{equation}
which holds true when the yardstick length $\delta $ is small enough to
measure accurately all boundary of the smallest cluster in sampling. When
the resolution is deficient, the Euclidean part of the perimeter length will
dominate the fractal one, so there is no way to find the fractal dimension
using the scaling relation of Eq.~(\ref{perima6}). It means that if the
length of a fractal curve was measured too roughly with the very large
yardstick, its fractal properties could not be detected, and therefore such
a geometrical object would be manifested itself as Euclidean one. It is just
the resolution deficiency of this kind occurs at the crossover point in Fig.~%
\ref{figure2}. Starting with the cluster area less than 0.023 $\mu $m$^{2}$
(corresponding to the currents of $i>6$) it is impossible to measure all ``
skerries'' and `` fjords'' on the cluster coastlines, whereas all the
clusters of area less than the size of the measuring cell ($3.6\times
10^{-3} $ $\mu $m$^{2}$ that relates to the currents of $i>23$), exhibit
themselves as objects of Euclidean boundaries $\left( D=1\right) $. This
resolution deficiency can be also observed in Fig.~\ref{figure1}: some
crosses at its lower left corner are arranged discretely with the spacing
equal to the limit of resolution (60\thinspace nm), because some marks for
smallest clusters coincide for the finite resolution of the picture
digitization procedure.

The coastline fractal dimension was found above by means of regression
analysis of the whole primary sampling, where the very small clusters of
sizes lying at the breaking point of the resolution limit were also
included. Therefore, it is necessary to control how much the estimated value
of the fractal dimension could be distorted by the presence of such small
clusters in that sampling. For this purpose the truncated sampling has been
formed in such a way that only 380 clusters of the area greater than 0.0269 $%
\mu $m$^{2}$, for which the resolution deficiency is not appeared, have been
selected from the primary sampling. The corresponding points are plotted as
open circles (2) in Fig.~\ref{figure1}, whereas the results of statistical
treatment of this truncated sampling are presented in the third column of
Table~\ref{table1}. The least squares estimation of these perimeter-area
data gives the value of coastline fractal dimension $D=1.47\pm 0.03$ with
correlation coefficient 0.869. The slope of the regression line for the
truncated sampling (dotted line (4) in Fig.~\ref{figure1}) is slightly
steeper than for the primary one (solid line (3)). It is natural, because
the presence in the primary sampling of small clusters, which exhibit
themselves as Euclidean ones at the given resolution, leads to underrating
the found magnitude of fractal dimension. Nevertheless, the values of
fractal dimensions found for both samplings, virtually do not differ within
the accuracy of the statistical estimation. This is due to the high
robustness of procedure of the fractal dimension estimation on a basis of
the scaling relation of Eq.~(\ref{scaling4}): all the points both for the
primary sampling and for the truncated one fall on the same straight line,
without any bends or breaks (see Fig.~\ref{figure1}). At the same time, it
is necessary to note, that the empirical distribution function approach (see
Fig.~\ref{figure2}) provides the most data-sensitive technique of estimating
the resolution capability required to study the fractal properties of
clusters.

\section{THE PINNING GAIN FOR THE MAGNETIC FLUX TRAPPED IN FRACTALLY BOUNDED
CLUSTERS OF A NORMAL PHASE}

The found cumulative probability function of Eq.~(\ref{exhyp5}) allows us to
fully describe the effect of the transport current on the trapped magnetic
flux. Using this function, the probability density $f(i)\equiv dF/di$ for
the critical current distribution can be readily derived: 
\begin{equation}
f\left( i\right) =\frac{2}{D}\left( \frac{2+D}{2}\right) ^{\frac{2}{D}+1}i^{-%
\frac{2}{D}-1}\exp \left[ -\left( \frac{2+D}{2}\right) ^{\frac{2}{D}+1}i^{-%
\frac{2}{D}}\right]  \label{densi7}
\end{equation}
This function is normalized to unity over all possible positive values of
critical current. The use of the exponential-hyperbolic critical current
distribution of Eq.~(\ref{exhyp5}) allows us to avoid the inevitable
uncertainty caused by truncation of non-physical negative values of
depinning currents, as it takes place, for example, in the case of a normal
distribution \cite{k32}-\cite{b34}.

The exponential-hyperbolic distribution of Eq.~(\ref{exhyp5}) has such an
important property: the function $F=F(i)$ is extremely ``flat`` in the
vicinity of the co-ordinate origin. It is easy to show that all its
derivatives are equal to zero at the point of $i=0$: 
\[
\frac{d^{k}}{di^{k}}F(0)=0\text{ \ \ \ \ \ \ \ \ for any value of }k 
\]

Therefore even the Taylor series expansion in the vicinity of the origin
converges to zero, instead of the quantity $F$ itself. This mathematical
feature has a clear physical meaning: so small transport current does not
affect the trapped magnetic flux, because there are no pinning centers of
such small critical currents in the overall statistical distribution, so
that all the vortices are still too strongly pinned to be broken away. As
can be seen from Fig.~\ref{figure2}, the change in the magnetic flux becomes
appreciable after the transition into a resistive state only (in the
transport current range of $i>1$).

The relative change in the trapped flux $\Delta \Phi /\Phi $, which can be
calculated from Eq.~(\ref{exhyp5}), also defines the density of vortices $n$
broken away from the pinning centers by the current $i$: 
\begin{equation}
n\left( i\right) =\frac{B}{\Phi _{0}}\int\limits_{0}^{i}f\left( i^{\prime
}\right) di^{\prime }=\frac{B}{\Phi _{0}}\frac{\Delta \Phi }{\Phi }
\label{volt8}
\end{equation}
where $B$ is the magnetic field, $\Phi _{0}\equiv hc/\left( 2e\right) $ is
the magnetic flux quantum ($h$ is Planck's constant, $c$ is the velocity of
light, $e$ is the electron charge). Figure~\ref{figure2} also displays such
a practically important property of superconducting structure containing
fractal clusters of a normal phase: the fractality intensifies the magnetic
flux trapping, hindering its breaking away from pinning centers, and thereby
enhances the critical current which sample is capable to withstand,
remaining in a superconducting state. Really, the transport current of a
magnitude $i=2$ causes the 43\% of the total trapped magnetic flux to break
away from the usual Euclidean clusters (curve (3)), whereas this value is
equal only to 25\% for fractal normal phase clusters of coastline dimension $%
D=1.44$ (curve (2)). It is equivalent to pinning reinforcement on 73\% in
the latter case. Thus the pinning amplification due to the fractality can be
characterized by the pinning gain factor 
\[
k_{D}\equiv \frac{\Delta \Phi \left( D=1\right) }{\Delta \Phi \left( \text{%
{\it current value of }D}\right) } 
\]
which is equal to relative decrease in the fraction of magnetic flux broken
away from fractal clusters of coastline dimension $D$ compared to the case
of Euclidean ones $\left( D=1\right) $. This quantity can be calculated from
the following formula: 
\[
k_{D}=\exp \left[ \left( \frac{2+D}{2}\right) ^{\frac{2}{D}+1}i^{-\frac{2}{D}%
}-\frac{3.375}{i^{2}}\right] 
\]

The characteristic dependencies of the pinning gain on a transport current
as well as on fractal dimension are given in Fig.~\ref{figure3}. The highest
amplification is reached when the cluster boundaries have the greatest
possible fractality: $%
\mathrel{\mathop{\max }\limits_{D}}%
k_{D}=k_{2}=\exp \left( (4i-3.375)/i^{2}\right) $, with the maximum of $%
k_{2} $ at transport current $i=1.6875$. Let us note that the pinning gain
characterizes the properties of a superconductor in the range of the
transport currents corresponding to a resistive state $\left( i>1\right) $.
At smaller current the total trapped flux remains unchanged (see Fig.~\ref
{figure2}) for lack of pinning centers of such small critical currents, so
the breaking of the vortices away has not started yet. When a transport
current becomes greater than the current of a resistive transition, some
finite resistance appears, so that the passage of electric current is
accompanied by the energy dissipation. As for any hard superconductor
(type-II, with pinning centers) this dissipation does not mean the
destruction of phase coherence yet. Some dissipation always accompanies any
motion of a magnetic flux that can happen in a hard superconductor even at
low transport current. Therefore the critical current in such materials
cannot be specified as the greatest non-dissipative current. The
superconducting state collapses only when a growth of dissipation becomes
avalanche-like as a result of thermo-magnetic instability.

The principal reason of pinning enhancement due to the fractality of the
normal phase clusters lies in the fundamental properties of the critical
current distribution. Figure~\ref{figure4} demonstrates the peculiarities of
the fractal probability density specified by Eq.~(\ref{densi7}). As in Fig.~%
\ref{figure1}, the thin lines show the extreme cases of Euclidean clusters $%
\left( D=1\right) $ and clusters of boundary with the maximum fractality $%
(D=2)$. As may be clearly seen from these graphs, the bell-shaped curve of
the distribution broadens out, moving towards greater magnitudes of current
as the fractal dimension increases. This shift can be described by
dependencies of average and mode of the critical current distribution on the
fractal dimension, as it is shown in the inset of Fig.~\ref{figure4}. The
mode of the distribution, which is equal to the value of the critical
current that provides the maximum of the probability density of Eq.~(\ref
{densi7}), depends linearly on the fractal dimension: {\it mode}$\,f\left(
i\right) =\left( 2+D\right) /2$. The average critical current obeys much
more strong superlinear law specified by Euler gamma function: 
\[
\overline{i}=\left( \frac{2+D}{2}\right) ^{\frac{2+D}{2}}\Gamma \left( 1-%
\frac{D}{2}\right) 
\]
The mean value of the critical current for Euclidean clusters is equal to $%
\overline{i}\left( D=1\right) =\left( 3/2\right) ^{3/2}\sqrt{\pi }=3.2562$,
while for clusters of maximum fractality this value becomes infinite: $%
\overline{i}\left( D=2\right) \rightarrow \infty $. Figure~\ref{figure4}
clearly demonstrates that increasing the fractal dimension gives a growth of
the contribution made by clusters of greater critical current to the overall
distribution, resulting just in enhancement of the magnetic flux trapping.

\section{ELECTRIC FIELD IN THE RESISTIVE STATE}

In the resistive state the hard superconductor is adequately specified by
its voltage-current ({\it V-I}) characteristic. The critical current
distribution of Eq.~(\ref{densi7}) allows us to find the electric field
arising from the magnetic flux motion after the vortices have been broken
away from the pinning centers. Inasmuch as each normal phase cluster
contributes to the total critical current distribution, the voltage across a
superconductor $V=V\left( i\right) $ is the response to the sum of effects
made by the contribution from each cluster . Such a response can be
expressed as a convolution integral: 
\begin{equation}
V=R_{f}\int\limits_{0}^{i}\left( i-i^{\prime }\right) f\left( i^{\prime
}\right) di^{\prime }  \label{volt9}
\end{equation}
where $R_{f}$ is the flux flow resistance. The similar approach is used
universally to consider behavior of the clusters of pinned vortex filaments 
\cite{w35}; to analyze the critical scaling of {\it V-I} characteristics of
superconductors \cite{b34}; that is to say, in all the cases where the
distribution of the depinning currents occurs. The present consideration
will be primarily concentrated on the consequences of the fractal nature of
the normal phase clusters specified by the distribution of Eq.~(\ref{densi7}%
), so all the problems related to possible dependence of the flux flow
resistance $R_{f}$ on a transport current will not be taken up here.

Let us consider the simplest case, wherein all of the pinning centers would
have an identical critical current $i_{c}$ so all the vortices would be
broken away simultaneously at $i=i_{c}$. Then, referring to Eq.~(\ref{volt8}%
), their density would have the following form: 
\[
n=\frac{B}{\Phi _{0}}\int\limits_{0}^{i}\delta \left( i^{\prime
}-i_{c}\right) di^{\prime }=\frac{B}{\Phi _{0}}h\left( i-i_{c}\right) 
\]

where $\delta \left( i\right) $ is Dirac delta function,

$h\left( i\right) \equiv \left\{ 
\begin{array}{cc}
1 & \text{\ \ \ for \ \ }i\geqslant 0 \\ 
0 & \text{ \ \ for \ \ }i<0
\end{array}
\right. $ \ \ \ is Heaviside step function.

Thus, the trapped flux would change by 100\% at once: $\Delta \Phi /\Phi
=h\left( i-i_{c}\right) $. Let us note that in this case $i_{c}=1$ due to
the convenient normalization chosen above: $i\equiv I/I_{c}$.

In the model case of delta-shaped distribution of the critical currents the
voltage across a superconductor in the flux flow regime, according to Eq.~(%
\ref{volt9}), would obey the simple linear law: $V=R_{f}\left(
i-i_{c}\right) h\left( i-i_{c}\right) $. The corresponding {\it V-I}
characteristic is shown in Fig.~\ref{figure5} by the dotted line (1).

For fractal distribution of critical currents the situation is quite
different, because the vortices are being broken away now in a wide range of
the transport currents. After the substitution of the function of Eq.~(\ref
{densi7}) in Eq.~(\ref{volt9}), upon integration by parts, the voltage
across a superconductor can be expressed with the cumulative probability
function of Eq.~(\ref{exhyp5}): 
\begin{equation}
V=R_{f}\int\limits_{0}^{i}F\left( i^{\prime }\right) di^{\prime }
\label{voltgen10}
\end{equation}
integration of which gives: 
\begin{equation}
V=R_{f}\exp \left[ -\left( \frac{2+D}{2}\right) ^{\frac{2}{D}+1}i^{-\frac{2}{%
D}}\right] \left\{ i-\left( \frac{2+D}{2}\right) ^{\frac{2+D}{2}}U\left[ 
\frac{D}{2},\frac{D}{2},\left( \frac{2+D}{2}\right) ^{\frac{2}{D}+1}i^{-%
\frac{2}{D}}\right] \right\}  \label{volthyp11}
\end{equation}
where $U\left( a,b,z\right) $ is Tricomi confluent hypergeometric function.

In extreme cases for $D=1$ and for $D=2$ expression of Eq.~(\ref{volthyp11})
can be simplified (see Appendix):

\begin{description}
\item  (a) Euclidean clusters $(D=1)$: 
\begin{equation}
V=R_{f}\left[ i\exp \left( -\frac{3.375}{i^{2}}\right) -\sqrt{3.375\pi }%
\text{erfc}\left( \frac{\sqrt{3.375}}{i}\right) \right]   \label{erf12}
\end{equation}

\item  where erfc$\left( z\right) $ is the complementary error function.

\item  (b) Clusters of boundary with the maximum fractality $\left(
D=2\right) $: 
\begin{equation}
V=R_{f}\left[ i\exp \left( -\frac{4}{i}\right) +4%
\mathop{\rm Ei}%
\left( -\frac{4}{i}\right) \right]   \label{ei13}
\end{equation}

\item  where $%
\mathop{\rm Ei}%
\left( z\right) $ is the exponential integral function.
\end{description}

The {\it V-I} characteristics of a superconductor containing fractal normal
phase clusters are presented in Fig.~\ref{figure5}. All the curves are
virtually starting with the transport current value of $i=1$ that is agreed
with the onset of the resistive state found above from cumulative
probability function of Eq.~(\ref{exhyp5}). When the current increases the
trapped flux remains unchanged until the vortices start to break away from
the pinning centers. As long as the magnetic flux does not move, no electric
field is arisen. Two thin lines (2), calculated using the formulas of Eq.~(%
\ref{erf12}) and Eq.~(\ref{ei13}), bound the region the {\it V-I}
characteristics can fall within for any possible values of fractal
dimension. As an example, the curve (3) demonstrates the {\it V-I}
characteristic of a superconductor containing fractal clusters of previously
obtained coastline dimension $D=1.44$. This figure shows that the fractality
reduces appreciably an electric field arising from the magnetic flux motion.
This effect is especially strong in the range of the currents $1<i<3$, where
the pinning enhancement also has a maximum (see Fig.~\ref{figure3}). Both
these effects have the same nature, inasmuch as their reason consists in the
peculiarities of fractal distribution of critical currents of Eq.~(\ref
{densi7}). As is seen from Fig.~\ref{figure4}, an increase of fractality
causes a significant broadening of the tail of the distribution $f=f\left(
i\right) $. It means that more and more of small clusters, which can best
trap the magnetic flux, are being involved in the game. Hence the density of
vortices broken away from pinning centers by the Lorentz force is reducing,
so the smaller part of a magnetic flux can flow, creating the smaller
electric field. In turn, the smaller the electric field is, the smaller is
the energy dissipated when the transport current passes through the sample.
Therefore, the decrease in heat-evolution, which could cause transition of a
superconductor into a normal state, means that the current-carrying
capability of the superconductor containing such fractal clusters is
enhanced.

Thus, Fig.~\ref{figure5}, as well as Fig.~\ref{figure3}, obviously
demonstrates such a practically important result: the fractality of the
boundary of the normal phase clusters, which act as the pinning centers,
prevents the destruction of a superconductivity by a transport current, and
therefore, causes the critical current to increase.

\section{CONCLUSION}

Thus, the fractal properties of the normal phase clusters have an essential
influence on the dynamics of the trapped magnetic flux. The crucial change
of the critical current distribution caused by increasing of the coastline
fractal dimension of the normal phase clusters forms the basis of this
effect. The most important result is that the fractality of cluster boundary
strengthens the flux pinning and thereby hinders the destruction of
superconductivity by the transport current, resulting in enhancement of the
current-carrying capability of a superconductor. This phenomenon provides
the principally new possibilities for increasing the critical current value
of composite superconductors by optimizing their geometric morphological
properties.

\appendix

\section{ELECTRIC VOLTAGE ACROSS A SUPERCONDUCTOR CAUSED BY THE MAGNETIC
FLUX MOTION IN EXTREME CASES OF EUCLIDEAN CLUSTERS AND CLUSTERS OF MAXIMUM
FRACTALITY}

In order to obtain the expressions of Eq.~(\ref{volthyp11}) - Eq.~(\ref{ei13}%
) for the voltage across a superconductor, it is necessary to integrate the
cumulative probability function for the critical current distribution of
Eq.~(\ref{exhyp5}). The substitution of exponential-hyperbolic distribution
of Eq.~(\ref{exhyp5}) in Eq.~(\ref{voltgen10}) gives: 
\begin{equation}
\frac{V}{R_{f}}=\int\limits_{0}^{i}dx\exp \left( -Cx^{-\frac{2}{D}}\right) 
\text{ \ \ \ \ , \ \ \ \ \ \ \ \ \ \ \ where \ \ \ \ \ \ }C\equiv \left( 
\frac{2+D}{2}\right) ^{\frac{2}{D}+1}  \label{a1}
\end{equation}
Using the change of a variable of the form $y\equiv Cx^{-2/D}$, we can get
the following expression: 
\[
\frac{V}{R_{f}}=\frac{D}{2}C^{\frac{D}{2}}\int\limits_{C\,i^{-\frac{2}{D}%
}}^{\infty }dye^{-y}y^{-\frac{2+D}{2}} 
\]
which, upon integration by parts, becomes: 
\begin{equation}
\frac{V}{R_{f}}=i\exp \left( -C\,i^{-\frac{2}{D}}\right) -C^{\frac{D}{2}%
}\Gamma \left( 1-\frac{D}{2},C\,i^{-\frac{2}{D}}\right)  \label{a2}
\end{equation}
where $\Gamma \left( \nu ,z\right) \equiv \int_{z}^{\infty }dye^{-y}y^{\nu
-1}$ is the incomplete gamma function. This function can be represented as: 
\begin{equation}
\Gamma \left( \nu ,z\right) =e^{-z}U\left( 1-\nu ,1-\nu ,z\right)  \label{a3}
\end{equation}
where $U\left( a,b,z\right) \equiv \left( \Gamma \left( a\right) \right)
^{-1}\int_{0}^{\infty }dye^{-zy}y^{a-1}\left( 1+y\right) ^{b-a-1}$ is
Tricomi confluent hypergeometric function; and $\Gamma \left( a\right)
\equiv \int_{0}^{\infty }dye^{-y}y^{a-1}$is Euler gamma function.

Thus with the help of Eq.~(\ref{a3}), the expression for the voltage across
a superconductor of Eq.~(\ref{a2}) can be written in its final form: 
\begin{equation}
\frac{V}{R_{f}}=\exp \left( -C\,i^{-\frac{2}{D}}\right) \left[ i-C^{\frac{D}{%
2}}U\left( \frac{D}{2},\frac{D}{2},C\,i^{-\frac{2}{D}}\right) \right]
\label{a4}
\end{equation}
This formula is similar to the expression of Eq.~(\ref{volthyp11}). The
corresponding {\it V-I} characteristic of a superconductor calculated using
this expression at $D=1.44$ is shown in Fig.~\ref{figure5} by the curve (3).

The equation (\ref{a4}) can be transformed to more simple form in two
special cases:

\begin{description}
\item  (a) For clusters of Euclidean boundary $\left( D=1\right) $:

\item  At $D=1$ the following representation is valid for Tricomi confluent
hypergeometric function: 
\[
U\left( \frac{1}{2},\frac{1}{2},z\right) =\sqrt{\pi }e^{z}\text{erfc}\left( 
\sqrt{z}\right) 
\]

\item  where erfc$\left( z\right) \equiv \left( 2/\sqrt{\pi }\right)
\int_{z}^{\infty }dye^{-y^{2}}$ is the complementary error function.

\item  The substitution of this representation into the equation (~\ref{a4})
gives the same expression for the voltage across a superconductor as the
formula of Eq.~(\ref{erf12}): 
\begin{equation}
\frac{V}{R_{f}}=i\exp \left( -\frac{C}{i^{2}}\right) -\sqrt{\pi C}\text{erfc}%
\left( \frac{\sqrt{C}}{i}\right)   \label{a5}
\end{equation}

\item  where, according to Eq.~(\ref{a1}), $C=3.375$.

\item  (b) For clusters of boundary with the maximum fractality $\left(
D=2\right) $:

\item  At $D=2$ there is such a representation for Tricomi confluent
hypergeometric function: 
\[
U\left( 1,1,z\right) =-e^{z}%
\mathop{\rm Ei}%
\left( -z\right) 
\]

\item  where $%
\mathop{\rm Ei}%
\left( -z\right) \equiv \int_{-\infty }^{-z}dy\frac{e^{y}}{y}$, $z>0$, is
the exponential integral function.

\item  Taking into account this formula, the expression (\ref{a4}) for the
voltage across a superconductor can be re-written as: 
\begin{equation}
\frac{V}{R_{f}}=i\exp \left( -\frac{C}{i}\right) +C%
\mathop{\rm Ei}%
\left( -\frac{C}{i}\right)   \label{a6}
\end{equation}

\item  where, according to Eq.~(\ref{a1}), $C=4$. The last formula coincides
with the expression of Eq.~(\ref{ei13}).
\end{description}

The formulas of Eqs.~(\ref{a5}), (\ref{a6}) describe dependencies of the
voltage across a superconductor in a resistive state on the transport
current for extreme values of the coastline fractal dimension. Two
corresponding {\it V-I} curves are shown in Fig.~\ref{figure5} by thin lines
(2). Whatever the geometric morphological properties of the normal phase
clusters may be, the {\it V-I} characteristics of a superconductor will fall
within the region bounded by those two limiting curves, as it is shown in
Fig.~\ref{figure5} (like the curve (3) drawn for $D=1.44$).

\begin{center}
\begin{table}[tbp] \centering%
%
\caption{Statistics of normal phase clusters and estimation of fractal
dimension\label{table1}} 
\begin{tabular}{ccc}
& Primary sampling & Truncated sampling \\ \hline\hline
Sampling size & $528$ & $380$ \\ \hline\hline
Mean $A$, $\mu $m$^{2}$ & $0.0765$ & $0.1002$ \\ 
Sample standard deviation of $A$, $\mu $m$^{2}$ & $0.0726$ & $0.0729$ \\ 
Standard error of estimate for $A,$ $\mu $m$^{2}$ & $3.16\times 10^{-3}$ & $%
3.74\times 10^{-3}$ \\ 
Total scanned $A,$ $\mu $m$^{2}$ & $40.415$ & $38.093$ \\ 
Min value of $A$, $\mu $m$^{2}$ & $2.07\times 10^{-3}$ & $0.0269$ \\ 
Max value of $A$, $\mu $m$^{2}$ & $0.4015$ & $0.4015$ \\ \hline\hline
Mean $P,\mu $m & $1.293$ & $1.616$ \\ 
Sample standard deviation of $P$, $\mu $m & $0.962$ & $0.949$ \\ 
Standard error of estimate for $P$, $\mu $m & $0.0419$ & $0.0487$ \\ 
Total scanned $P$, $\mu $m & $682.87$ & $614.19$ \\ 
Min value of $P$, $\mu $m & $0.096$ & $0.515$ \\ 
Max value of $P$, $\mu $m & $5.791$ & $5.791$ \\ \hline\hline
Correlation coefficient & $0.929$ & $0.869$ \\ 
Estimated fractal dimension $D$ & $1.44$ & $1.47$ \\ 
Standard deviation of $D$ & $0.02$ & $0.03$%
\end{tabular}
\end{table}%
%
\end{center}

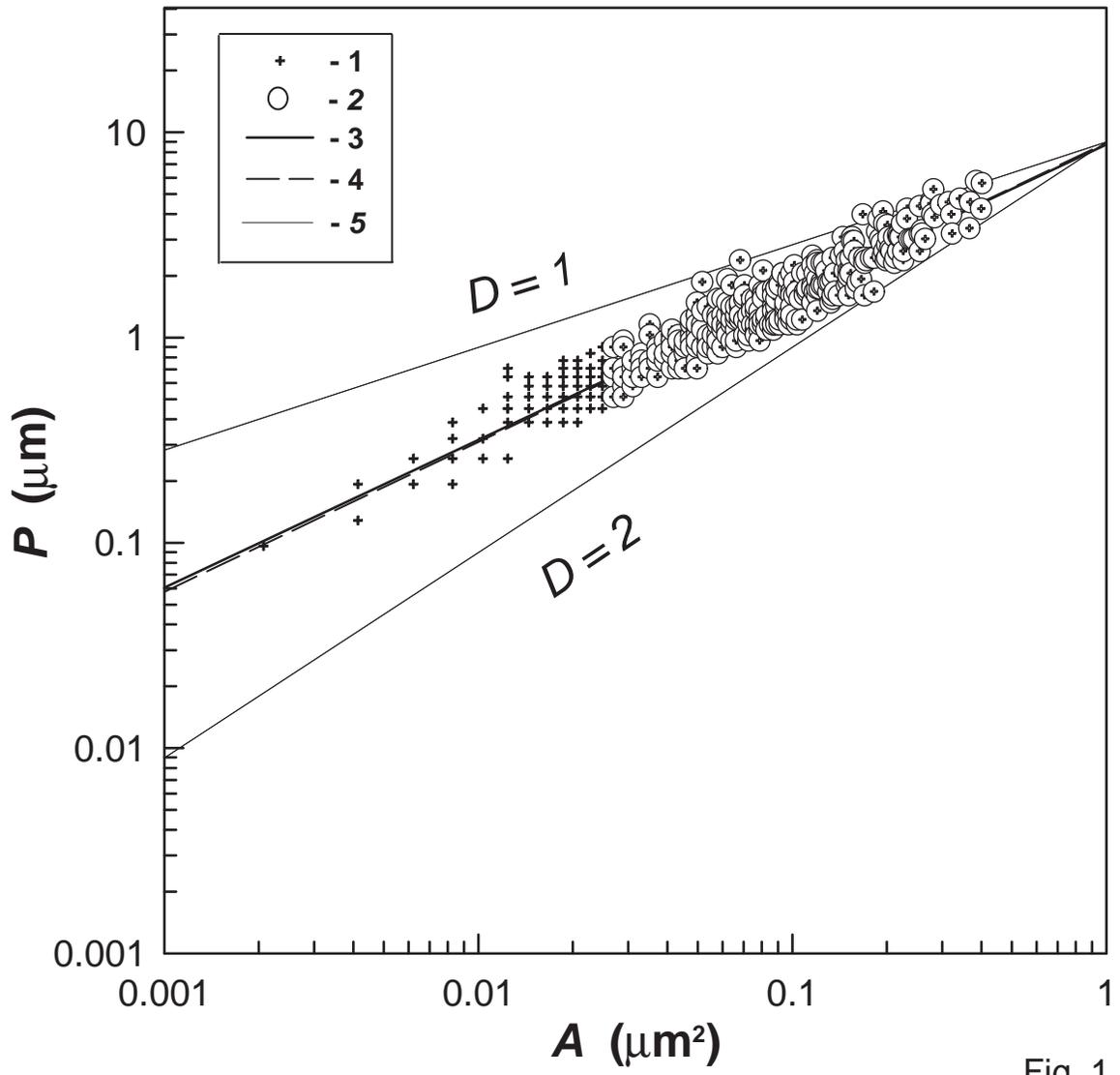
\begin{figure}[tbp]
\caption{Perimeter-area relationship for the normal phase clusters with
fractal boundary. Plot (1) shows the data of the primary sampling (528
points); plot (2) shows the data of the truncated sampling (380 points);
line (3) is the least square regression line for the primary sampling; line
(4) is the least square regression line for the truncated sampling. Two
lines (5) display the range of slope that the perimeter-area curves can have
for any possible fractal dimension $D$. ($D=1$ - for clusters of Euclidean
boundary, $D=2$ - for clusters of boundary with the maximum fractality).}
\label{figure1}
\end{figure}
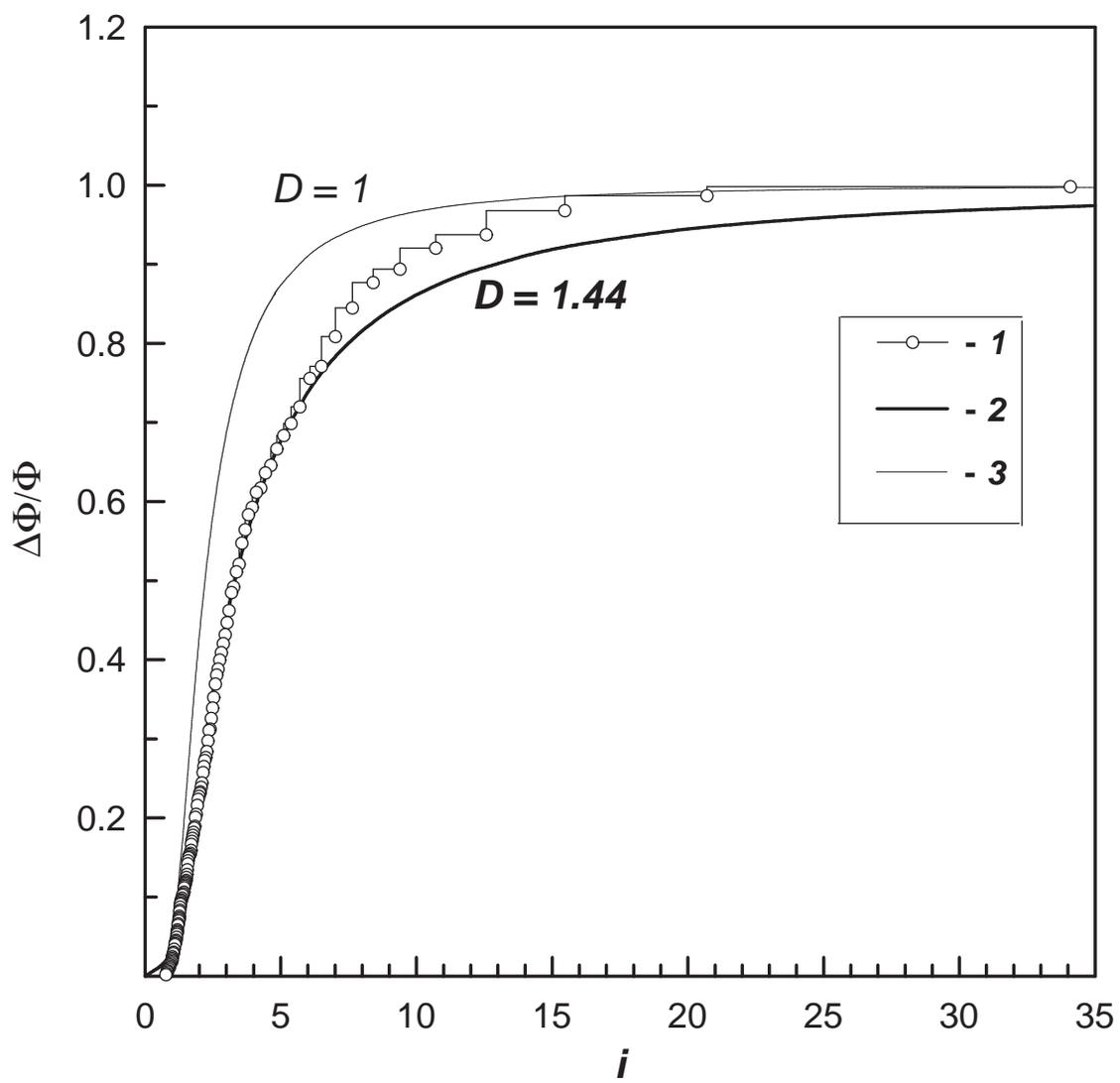
\begin{figure}[tbp]
\caption{Effect of a transport current on the magnetic flux trapped in
fractal clusters of a normal phase. Step line with open circles (1) is the
sample empirical function of critical current distribution; line (2) shows
the decrease in trapped flux for the fractal clusters of coastline dimension 
$D=1.44$; line (3) shows the decrease in trapped flux for Euclidean clusters
of coastline dimension $D=1$.}
\label{figure2}
\end{figure}
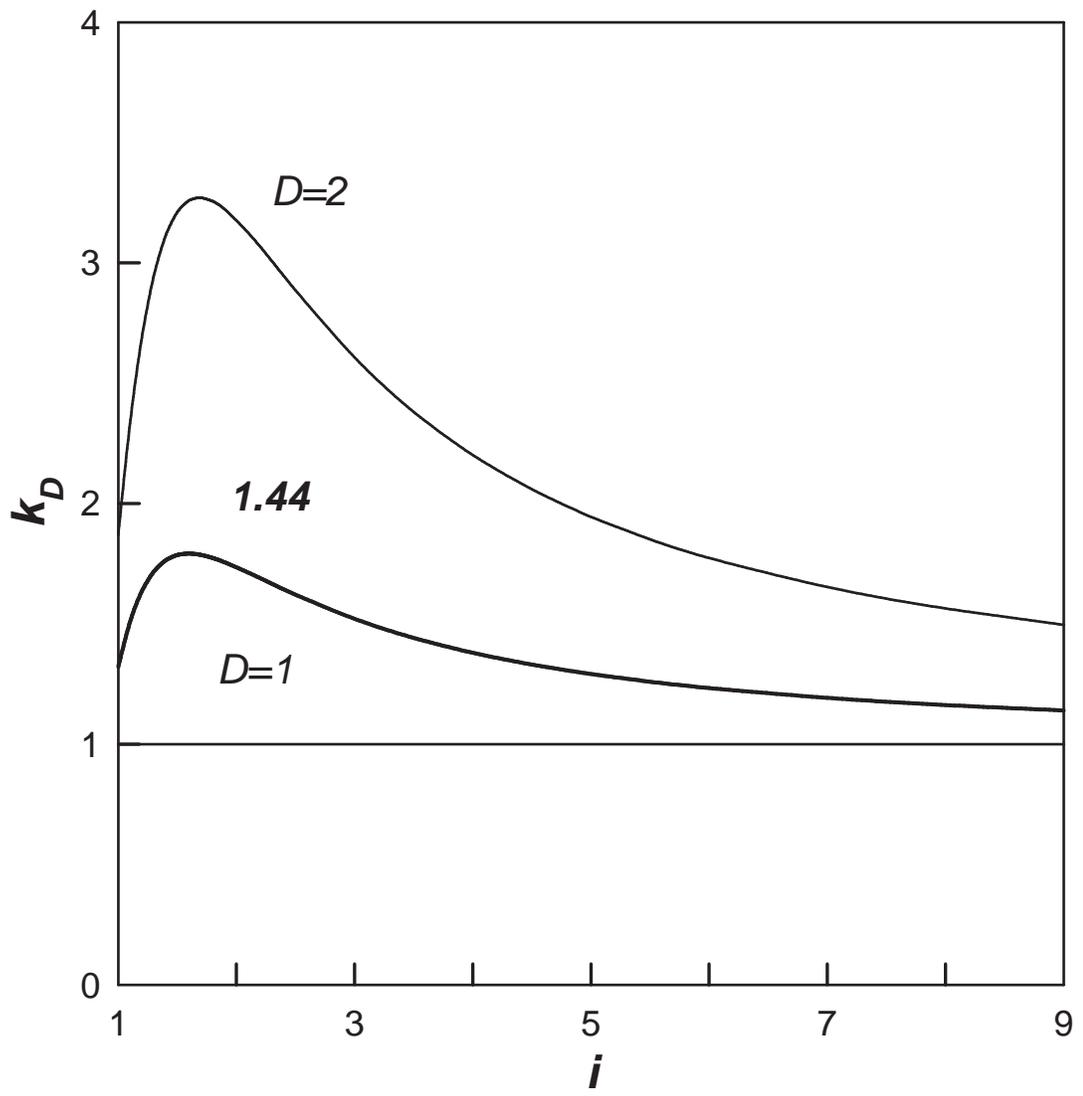
\begin{figure}[tbp]
\caption{Pinning gain for an arbitrary coastline fractal dimension of the
cluster perimeter.}
\label{figure3}
\end{figure}
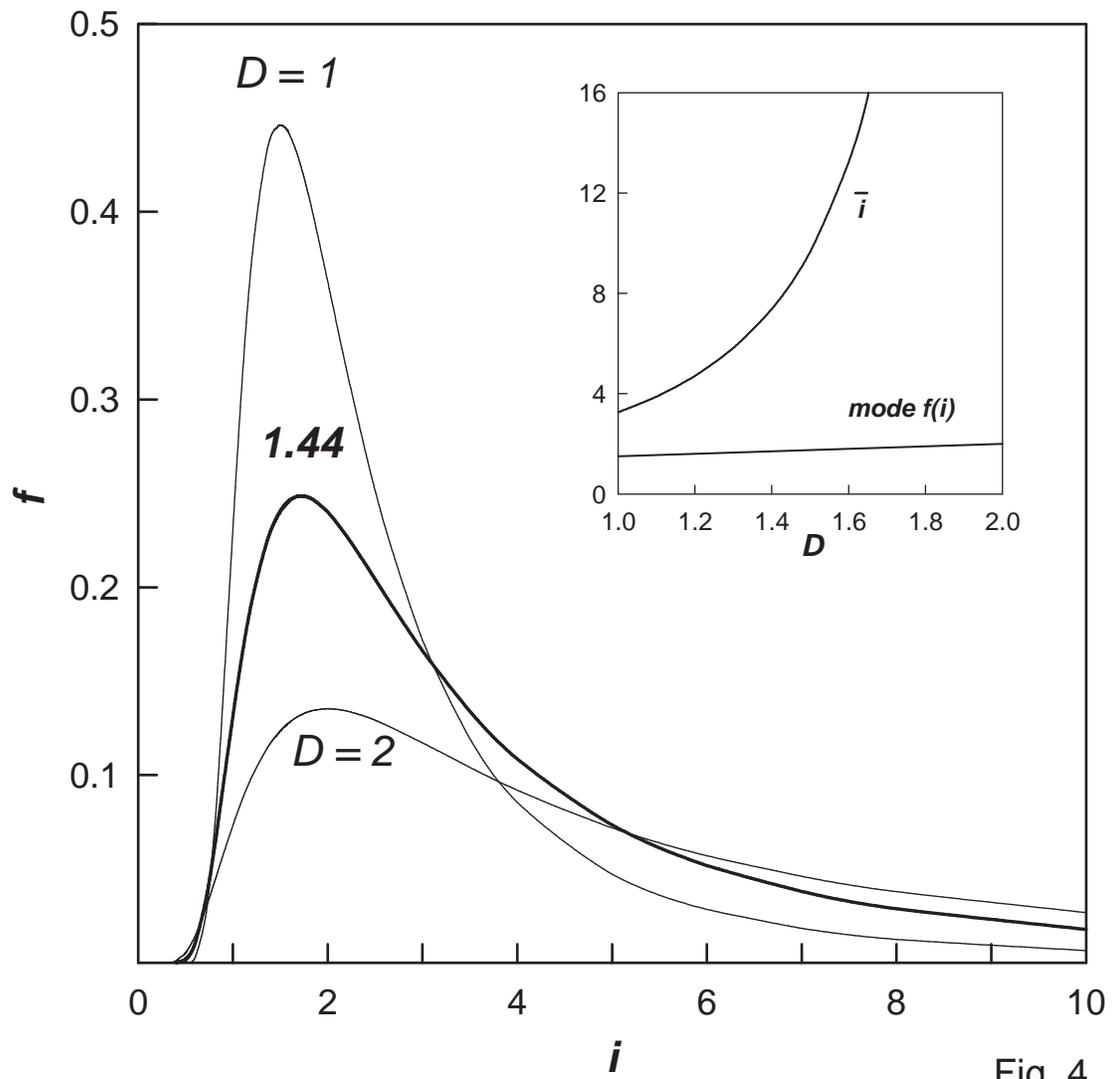
\begin{figure}[tbp]
\caption{Influence of the fractal dimension of the perimeter of the normal
phase clusters on the critical current distribution. The inset shows the
dependencies of the average critical current $\overline{i}$ and mode {\it %
mode\thinspace }$f\left( i\right) $ of this distribution on the coastline
fractal dimension.}
\label{figure4}
\end{figure}
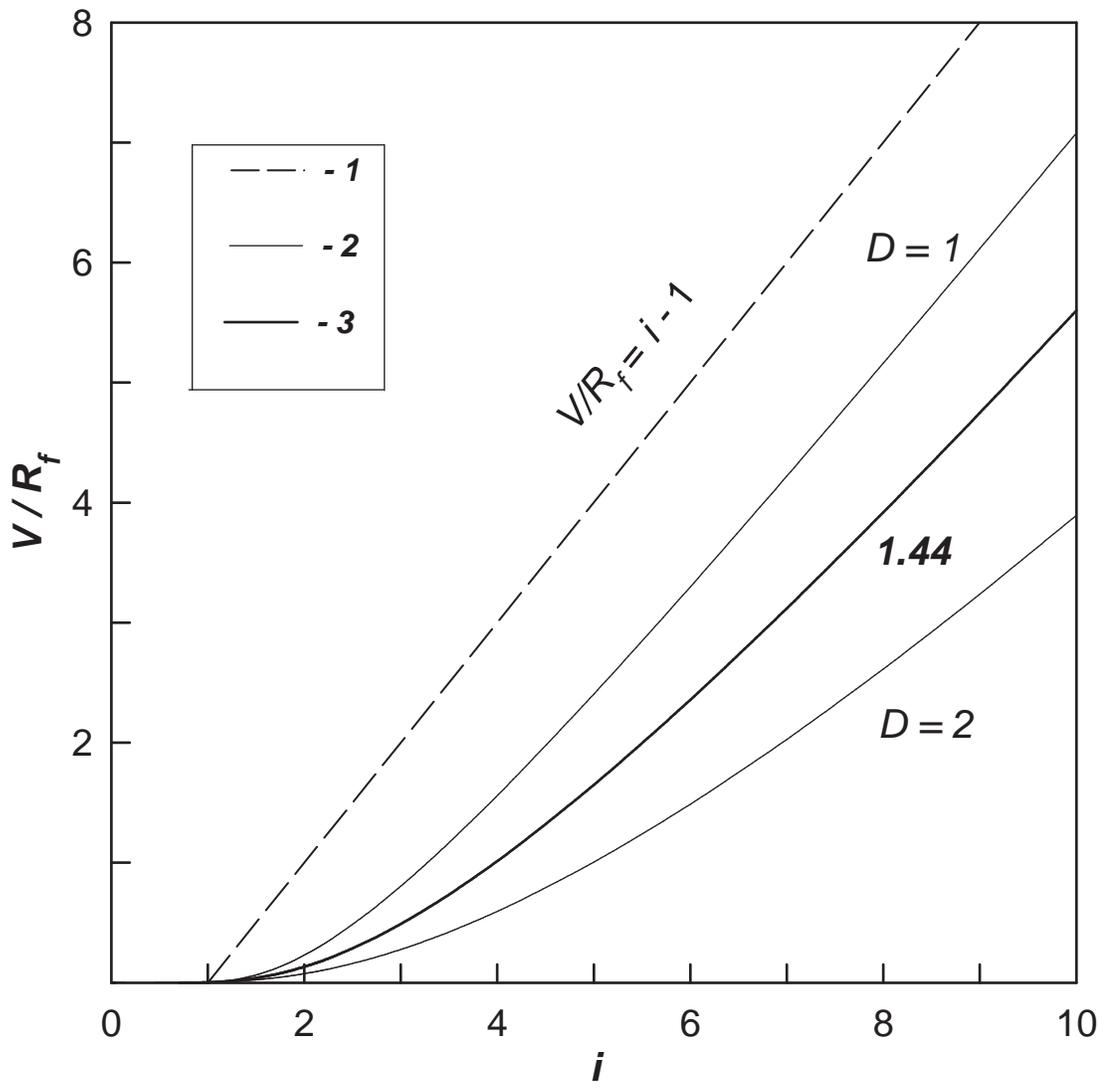
\begin{figure}[tbp]
\caption{Voltage-current characteristics of superconductors containing
fractal clusters of a normal phase. Dotted line (1) corresponds to the
delta-shaped distribution of the critical currents; lines (2) - to extreme
dependencies of the voltage across a superconductor on the transport current
in the case of Euclidean clusters $\left( D=1\right) $ and clusters of
boundary with the maximum fractality $(D=2)$; line (3) - to {\it V-I}
characteristic of superconductor containing the normal phase clusters of
fractal dimension of the perimeter $D=1.44$.}
\label{figure5}
\end{figure}

\end{document}